\documentclass[a4paper,conference]{IEEEtran}

\usepackage{ifpdf}

\usepackage{cite}

\ifCLASSINFOpdf
   \usepackage[pdftex]{graphicx}
   \graphicspath{{./graphics/}}
   \DeclareGraphicsExtensions{.pdf,.jpeg,.png}
\else
   \usepackage[dvips]{graphicx}
   \graphicspath{{./graphics/}}
   \DeclareGraphicsExtensions{.eps}
\fi

\usepackage{amsmath}

\usepackage{array}

\usepackage[caption=false,font=footnotesize]{subfig}
\usepackage{multirow}

\newcommand{\va}[1]{\vm{a}_{#1}}          
\newcommand{\vafirst}[1]{\vm{a'}_{#1}}          
\newcommand{\vasecond}[1]{\vm{a''}_{#1}}          

\newcommand{\VA}[1]{\mathcal{A}_{#1}}     
\newcommand{\Z}[1]{\mathcal{Z}_{#1}}               
\newcommand{\D}[1]{\mathcal{D}_{#1}}               
\newcommand{\C}[1]{\mathcal{C}_{#1}}               
\newcommand{\cutoff}{d_{\mathrm{c}}}             

\newcommand{\vu}[2]{\mbox{$#1\,\text{#2}$}} 
\newcommand{\vm}[1]{\mathbf{#1}}

\newcommand{\Tp}{{T}_p}
\newcommand{\sigmaa}{\sigma_a^2}
\newcommand{\sigmaz}{\sigma_z^2}

\newcommand{\tauLOSMLi}{\hat{\tau}_{\text{LOS}, \ell}^{(\text{ML}, i)}}
\newcommand{\tauLOSML}{\hat{\tau}_{\text{LOS}, \ell}^{(\text{ML})}}
\newcommand{\tauLOSJBSF}{\hat{\tau}_{\text{LOS}, \ell}^{(\text{JBSF})}}
\newcommand{\tauLOSJBSFi}{\hat{\tau}_{\text{LOS}, \ell}^{(\text{JBSF}, i)}}
\newcommand{\tausb}{\tau_{\text{sb}}}
\newcommand{\taumax}{\tau_{\text{max}}}
\newcommand{\vxmax}{v_{x,\text{max}}}

\begin{document}

\title{Accurate and Robust Indoor Localization \\ Systems using Ultra-wideband Signals}
\author{\IEEEauthorblockN{Paul Meissner$^*$, Erik Leitinger$^*$, Markus Fr\"ohle$^*$, and Klaus Witrisal$^*$}

\IEEEauthorblockA{$^*$Graz University of Technology, Austria; E-mail: paul.meissner@tugraz.at}
}

\maketitle

\begin{abstract}
Indoor localization systems that are accurate and robust with respect to propagation channel conditions are still a technical challenge today. In particular, for systems based on range measurements from radio signals, non-line-of-sight (NLOS) situations can result in large position errors. In this paper, we address these issues using measurements in a representative indoor environment. Results show that conventional tracking schemes using high- and a low-complexity ranging algorithms are strongly impaired by NLOS conditions unless a very large signal bandwidth is used. Furthermore, we discuss and evaluate the performance of multipath-assisted indoor navigation and tracking (MINT), that can overcome these impairments by \emph{making use} of multipath propagation. Across a wide range of bandwidths, MINT shows superior performance compared to conventional schemes, and virtually no degradation in its robustness due to NLOS conditions.
\end{abstract}

\section{Introduction}

Indoor localization systems face many difficulties when they are to be deployed in practice \cite{Mautz2009}. Optical systems might become unusable in emergency situations due to fire and smoke; signal strength based systems are sensitive to fading effects; fingerprinting systems need dedicated training phases; etc. Range-measurement-based systems using radio signals, which are in the focus of this paper, offer advantages like the capability of being integrated in existing radio devices such as smartphones, the potential of low-power implementations and moderate infrastructure requirements. However, they face challenging performance impairments caused by propagation effects like strong reflections or diffuse scattering.

\begin{figure*}[t]
\centering
\includegraphics[width=.95\textwidth,keepaspectratio=true]{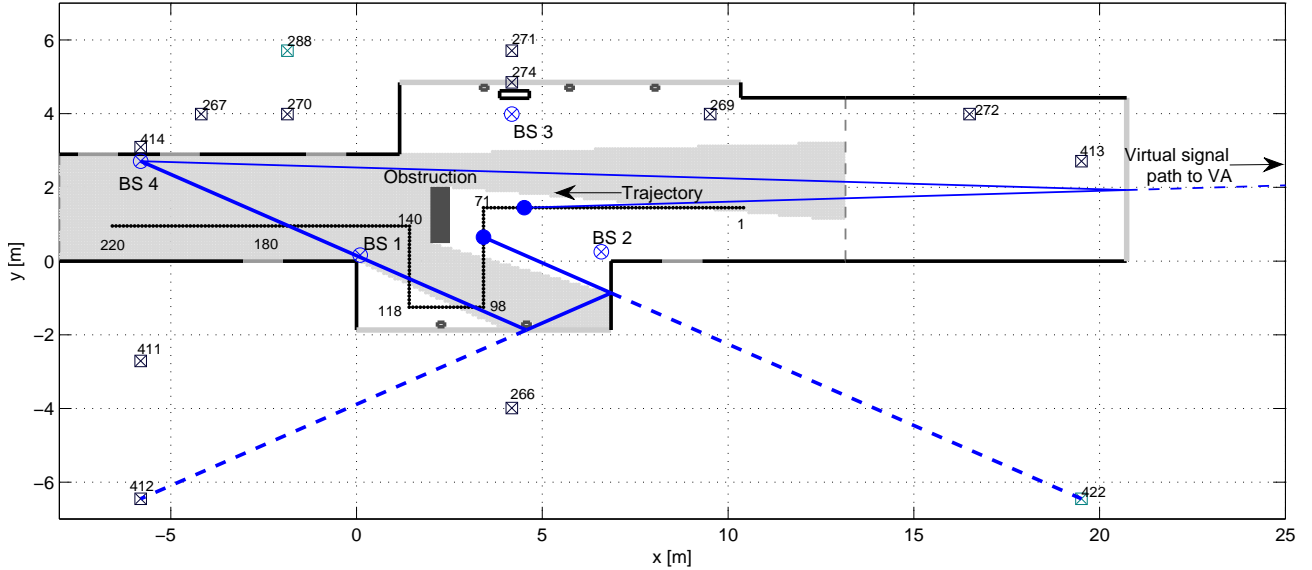}
\caption{Indoor tracking scenario considered in this paper. The black dotted line illustrates the user's trajectory in the scenario, selected position indices are given to allow easier interpretation of later results. Blue circled crosses are the base stations, while black and grey squares with crosses denote some few illustrative VAs for BSs 3 and 4. The area shaded in light gray denotes the geometric visibility region of the LOS path of BS 4, illustrating the NLOS region caused by the hypothetical obstruction in the center that is used for performance evaluations. Additionally, the thin and bold blue lines show two example propagation paths that can be modeled by VAs and that are potentially visible in the NLOS region.}
\label{fig:scen}
\end{figure*}

To counter these impairments, a large signal bandwidth is beneficial for two main reasons: First, the temporal resolution and therefore the attainable accuracy is increased. Second, the line-of-sight (LOS) path can be separated from other signal parts easier, which improves the robustness, i.e. the percentage of cases in which the attainable accuracy can be achieved. For these reasons, systems using ultra-wideband (UWB) signals are promising candidates for both accurate and robust indoor localization \cite{Gezici2009}.

In this paper, we examine the performance of a conventional range-based UWB indoor localization and tracking system in a representative indoor environment with several anchor nodes. Using channel measurements along a reference trajectory, mimicking the walking path of a pedestrian user, we show the dependence of the robustness and accuracy of the localization on changing propagation conditions. The LOS path is estimated using the maximum-likelihood method as well as using a low-complexity ranging algorithm. These estimates are provided as inputs to an extended Kalman filter (EKF), which tracks the user's position. The presence of strong reflections in the measurement signal can lead to a large bias in the estimate of the propagation delay of the direct path between user and anchor. This is especially true in non-LOS (NLOS) situations, where the direct path is blocked. Overall, a degradation of the robustness of the system is the result.

Usual countermeasures are to detect such situations based on signal statistics and discard the corresponding measurements \cite{Borras1998,Gezici2003}. In advanced schemes like \cite{Wymeersch2012,Marano2010}, machine learning techniques are used to mitigate the range bias in NLOS situations. However, these approaches discard position-related information that is inherent in the geometric structure of reflected MPCs. The usefulness of this information has been shown in \cite{Shen2009,Shen2010,WitrisalICC2012,MeissnerEUCAP2012}.

We have shown in previous work \cite{WitrisalICC2012,MeissnerIWSSIP2012,MeissnerEUCAP2012} that by using additional floor plan information, these reflected signal paths can be used for localization. Extracting these paths from the channel measurements and associating them to the floor plan, signal reflections can be seen as direct signals coming from virtual anchor (VA) nodes. In this way, one physically existing anchor node is turned into a set of virtual anchors, and NLOS situations no longer render this anchor useless for localization. This scheme, which we call multipath-assisted indoor navigation and tracking (MINT), can thus deliver the desired robustness with respect to the propagation conditions. However, the effect of diffuse scattering is especially pronounced in indoor environments due to the presence of objects like, e.g., furniture. This makes the extraction of the reflected components from the measurements and the required data association challenging. To highlight these issues in a practical setting, this paper extends our previous work in \cite{MeissnerIWSSIP2012} with the use of multiple base stations, a detailed performance analysis of the proposed algorithms and a comparison to standard tracking schemes employed in indoor localization.

The key contributions of this work are:
\begin{itemize}
  \item We show the performance and the robustness issues of a conventional, EKF based tracking system with respect to channel conditions. This is done in a representative indoor environment using measured signals.
  \item Making explicit use of multipath, robustness and accuracy can be increased. We discuss an implementation of the proposed MINT approach and examine its performance. Results show the excellent robustness of MINT with respect to NLOS situations.
  \item The results obtained highlight general practical considerations for indoor localization systems, i.e. the importance of a large bandwidth and propagation channel conditions is shown.
\end{itemize}

This paper is organized as follows: Section \ref{sec:scen} explains the scenario, the measurement campaign and the geometrical modeling of the environment. A short overview over performance bounds for MINT is given in Section \ref{sec:CRLB}, while Section \ref{sec:ranging} discusses channel estimation and ranging algorithms. Tracking is explained in Section \ref{sec:tracking} and a detailed discussion of results is given in Section \ref{sec:results}.

\section{Scenario, Measurements, and Geometric Modeling}
\label{sec:scen}

Fig. \ref{fig:scen} shows the scenario used in this paper. It is the ground floor of a large three-storey building with open ceilings in the corridor areas in all but the uppermost floors. Walls are made of reinforced concrete (shown as black outer lines) and doors (shown in grey) are made of metal. The long grey outer lines on the upper and lower side are large windows with some metal pillars in-between. We placed four base stations (BSs) at known locations and measured the UWB channel between them and a moving agent. The latter was moved along a trajectory consisting of $220$ points spaced by \vu{10}{cm}. The obstruction shown near the center of the corridor has not been present in the measurements. It is introduced artificially in the performance evaluations to show the influence of NLOS scenarios on the localization, c.f. Section \ref{sec:results}.

\subsection{Geometric Modeling of the Environment}
If a floor plan of the building is available, the BS positions $\va{i}$, $i=1,\dots, 4$, can be mirrored with respect to reflecting surfaces like walls to obtain so-called first-order virtual anchors (VAs) at positions $\vafirst{i}$. Using these, the procedure can be repeated to obtain second-order VAs $\vasecond{i}$, and so on. The  visibility regions of these VAs, i.e. the regions where the corresponding reflections are possible, can be computed by ray tracing as explained in \cite{MeissnerICUWB2011}. If the corresponding MPCs are detectable at the moving agent, they can be associated to the VAs and used for localization. Fig. \ref{fig:scen} shows a subset of VAs for BSs 3 and 4 together with two example signal paths.

In an analogous manner to \cite{MeissnerIWSSIP2012}, the overall set of VAs for the $i$-th BS is denoted as
\begin{equation}
  \VA{}^{(i)} = \{ \va{j} \; : \; j=1, \dots, N_{\mathrm{VA}}^{(i)} \}.
  \label{eq:allVAs}
\end{equation}
%
Here, the superscript indicating the VA order has been dropped as it is not important in the upcoming discussion. A function $f_\mathrm{vis}(\va{j}, \vm{p})$ determines the visibility of the $j$-th VA at a position $\vm{p}$, i.e.
\begin{equation}
  f_\mathrm{vis}(\va{j}, \vm{p}) = \begin{cases}
                                                   1, & \text{if VA } \va{j} \text{ is visible at } \vm{p}\\
                                                   0, & \text{else.}
                                                 \end{cases}
                                                 \label{eq:fvis}
\end{equation}
The set of expected visible VAs at a specific position $\vm{p}_\ell$ can then be computed evaluating \eqref{eq:fvis} for all VAs in \eqref{eq:allVAs}
\begin{equation}
  \VA{\ell}^{(i)} =\{ \va{\ell,1}, \dots, \va{\ell,K_\ell^{(i)}}\} = \{ \va{j} \; : \; f_\mathrm{vis}(\va{j}, \vm{p}_\ell)=1  \}.
  \label{eq:VAsl}
\end{equation}
In the following, we omit the BS index $i$ wherever it should be clear from the context that the respective quantity exists for all BSs.

It can be seen from \eqref{eq:VAsl} that a set of anchor nodes significantly larger than the set of BSs can potentially used for localization. We call this approach \emph{multipath-assisted indoor navigation and tracking} (MINT). However, the challenges are as manifold as the potential gains: First, the MPCs corresponding to the VAs need to  be detected in the signal received at the agent, which depends on influences such as e.g. material parameters or reflection angles. Second, together with deterministic MPCs, also some scattered, non-specular components will cause a certain level of clutter in the measurements. Third, the detected MPCs need to be associated to the VAs to make their position-related information accessible.
Summarizing, a moving agent needs to perform detection of MPCs in the presence of possibly strong diffuse multipath and noise as well as subsequently a robust data association procedure to infer its location.

\subsection{Channel Measurements}

\begin{figure}[t]
\centering
\includegraphics[width=1\columnwidth,keepaspectratio=true]{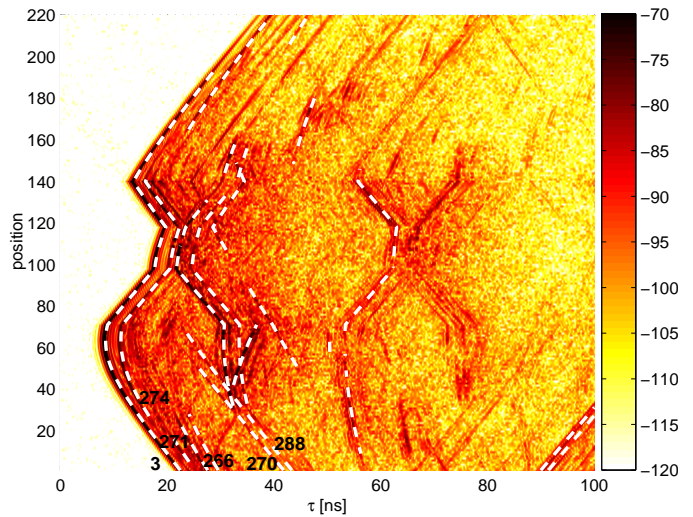}
\caption{Received signals $r_\ell(t)$ in logarithmic scaling over the trajectory for BS 3. The white dashed lines indicate expected delay paths of MPCs modeled by VAs up to order two, computed by geometric ray-tracing. Indices of some of the VAs shown in Fig. \ref{fig:scen} are given.}
\label{fig:hktau}
\end{figure}

Measurements were performed in the frequency domain with a vector network analyzer (VNA) over the full FCC bandwidth from $3.1$ to \vu{10.6}{GHz}. These measurements have also been used in \cite{MeissnerICUWB2011,MeissnerEUCAP2012,MeissnerIWSSIP2012}. The discrete frequency representation of the channel's transfer function at position $\vm{p}_\ell$ is denoted as $H_\ell[k]$. Using a frequency spacing of $\Delta f$, it corresponds to a Fourier series (FS) of a time-domain channel impulse response (CIR) $h_\ell(\tau)$. This CIR is periodic with a period of $\tau_{\text{max}}=1/\Delta f$, which is the maximum resolvable delay. The FS representation allows an arbitrarily fine time resolution $\Delta \tau$ to be achieved. Using an IFFT with size $N_{\text{FFT}}=\lceil (\Delta f \Delta \tau)^{-1} \rceil$, the time domain equivalent baseband signal is obtained as
\begin{equation}
  r_\ell(t) = \text{IFFT}_{N_{\text{FFT}}} \left\{ H_\ell[k] S[k]  \right\} \text{e}^{-j2\pi (f_c-f_0) t}.
  \label{eq:rlIFFT}
\end{equation}
In \eqref{eq:rlIFFT}, $S[k]$ is the discrete frequency domain representation of a raised-cosine pulse $s(t)$ with pulse duration $\Tp$ and roll-off factor $\beta_\text{R}$ \cite{Proakis2008}. It is used to cut out the desired band of the signal as well as to filter it to facilitate an MPC estimation algorithm. The frequencies $f_c$ and $f_0$ are the desired center frequency and the lower band edge of the extracted band, respectively.
This procedure is similar to \cite{Santos2010}. Examples for $r_\ell(\tau)$ obtained in this way are shown in Fig. \ref{fig:hktau} for BS 3 over the trajectory.

\section{Channel Influence on Localization}
\label{sec:CRLB}

The received signal at the $\ell$-th trajectory position $\vm{p}_\ell$, corresponding to the processing in \eqref{eq:rlIFFT}, is modeled as \cite{WitrisalICC2012}
\begin{equation}
  r_\ell(t) = \sum_{k=1}^{K_\ell} \alpha_{k,\ell} s(t-\tau_{k,\ell}) + s(t)*\nu_\ell(t) + w_\ell(t).
 \label{eq:rx_signal}
\end{equation}
%
It consists of a sum of $K_\ell$ deterministic MPCs, scaled and shifted by the complex amplitudes $\alpha_{k,\ell}$ and propagation delays $\tau_{k,\ell}$, respectively. The signal $\nu_\ell(t)$ is a stochastic process modeling the diffuse multipath (DM), i.e. diffuse scattering at rough surfaces and other propagation mechanisms that are not captured by the specular part of the channel. It is observed at the agent as a convolution with the transmit pulse $s(t)$. Finally, $w_\ell(t)$ denotes additive white Gaussian noise (AWGN) with two-sided power spectral density $N_0/2$.

We expect that the deterministic part of $r_\ell(t)$ can be modeled by the set of VAs $\VA{\ell}$. In \cite{WitrisalICC2012}, we have derived the Cram\'er-Rao lower bound (CRLB) on the position error for the MINT scheme, based on the received signal $r_\ell(t)$. It is given as the inverse of the equivalent Fisher Information matrix (EFIM) \cite{Shen2010}. If all $K_\ell$ deterministic MPCs are orthogonal at $\vm{p}_\ell$, i.e. there is no path overlap in the delay domain, the EFIM can be decomposed as
\begin{equation}
  \vm{J}_{\vm{p}_\ell} = \frac{8 \pi^2 \beta^2}{c^2} \sum_{k=1}^{K_\ell} \text{SINR}_{k,\ell} \vm{J}_\vm{r}(\phi_{k,\ell})
\label{eq:efim1}
\end{equation}
where $\beta$ denotes the effective (RMS) bandwidth of $s(t)$ and $c$ is the speed of light. The decomposition is instructive since it separates the channel's influence from the geometry of the VAs.
The former is given by the signal-to-interference-plus-noise-ratio (SINR) of the deterministic MPCs. For the $k$-th MPC at position $\vm{p}_\ell$, it is defined as \cite{WitrisalICC2012}
\begin{equation}
 \text{SINR}_{k,\ell} = \frac{|\alpha_{k,\ell}|^2}{N_0+ \Tp S_{\nu,\ell}(\tau_{k,\ell}) } 
 \label{eq:sinr}.
\end{equation}
Here, interference is understood as the contribution of the DM, which is given by the power delay profile (PDP) of the DM, $S_{\nu,\ell}$ at the delay $\tau_{k,\ell}$ of the respective MPC, scaled with the pulse duration. This PDP is defined as the second central moment of the DM process $\nu_\ell(t)$.

The influence of the geometry is accounted for by the ranging direction matrix $\vm{J}_\vm{r}(\phi_{k,\ell})$ \cite{Shen2010,WitrisalICC2012}, which is defined as
\begin{equation}
  \vm{J}_\vm{r}(\phi_{k,\ell}) = \begin{bmatrix}
                                \cos^2(\phi_{k,\ell}) & \cos(\phi_{k,\ell})\sin(\phi_{k,\ell})\\
                                \cos(\phi_{k,\ell})\sin(\phi_{k,\ell}) & \sin^2(\phi_{k,\ell})
                              \end{bmatrix}.
  \label{eq:RDM}
\end{equation}
This matrix has one eigenvector pointing in the direction $\phi_{k,\ell}$, the angle from the VA at $\vm{a}_k$ to position $\vm{p}_\ell$. The SINR in \eqref{eq:sinr} is used in the performance evaluations to emulate changes of propagation conditions caused by the artificial obstruction that is shown in Fig. \ref{fig:scen}.

\section{Channel Estimation and Ranging}
\label{sec:ranging}

Conventional time-of-arrival (ToA) localization requires an estimate of the time-of-flight of the signal from the BS to the agent. This process is called ranging \cite{Sahinoglu2008}. In this work, we consider two different ranging algorithms, maximum likelihood (ML) and jump-back search-forward (JBSF). The former is computationally intensive since it requires channel estimation, while the latter is a popular low-complexity algorithm.
The MINT approach requires range estimates to all the MPCs that are modeled by VAs, i.e., an estimation of the deterministic part of the UWB channel as for ML ranging.

\subsection{Channel Estimation and ML Ranging}

It has been shown that the MPCs corresponding to the VAs can carry a large fraction of the energy of the received signal \cite{MeissnerICUWB2011}, hence we extract those $\hat{K}_\ell$ MPCs that result in the minimum energy of the difference signal
\begin{equation}
  \{ \hat{\tau}_{k,\ell} \} = \arg \min_{ \{ \tau_{k,\ell} \} } \int_T \left|  r_\ell(t) - \sum_{k=1}^{\hat{K}_\ell} \hat{\alpha}_{k,\ell} s(t-\hat{\tau}_{k,\ell}) \right|^2 dt
  \label{eq:clean}
\end{equation}
where $T$ is the observation interval. Of course the parameter $\hat{K}_\ell$ should be related to $K_\ell$ in \eqref{eq:rx_signal}. However, without the knowledge of the position, a value has to be chosen that allows for the extraction of the geometrically relevant paths over the entire scenario. The choice of this parameter will be discussed in Section \ref{sec:results_param}.

For this work, we assume a separable channel, i.e. there is no overlap between deterministic MPCs. In this case, \eqref{eq:clean} can be performed path-per-path. The amplitudes $\alpha_{k,\ell}$ are nuisance parameters to this estimation and can be obtained by a projection of the unit-energy pulse $s(t)$, shifted to the estimated delay $\hat{\tau}_{k,\ell}$, onto the received signal
\begin{equation}
  \hat{\alpha}_{k,\ell} = \int_T r_\ell^*(t) s(t-\hat{\tau}_{k,\ell}) dt
  \label{eq:alpha}
\end{equation}
where $^*$ denotes complex conjugation. With the separable channel assumption, this procedure is similar to \cite{Win2002}.

For MINT, the set of estimated arrival times in \eqref{eq:clean}, converted to distances, constitutes the measurement input to the tracking algorithms. Additionally, a data association (DA) scheme is necessary, which will be described in Section \ref{sec:MINTDA}. For ML ranging, only the propagation delay of the first arriving path, the LOS path, is interesting and can be obtained from \eqref{eq:clean} as
\begin{equation}
  \tauLOSML = \min \{ \hat{\tau}_{k,\ell}  \}.
  \label{eq:MLrange}
\end{equation}

\begin{figure}[t]
\centering
\includegraphics[width=1\columnwidth,keepaspectratio=true]{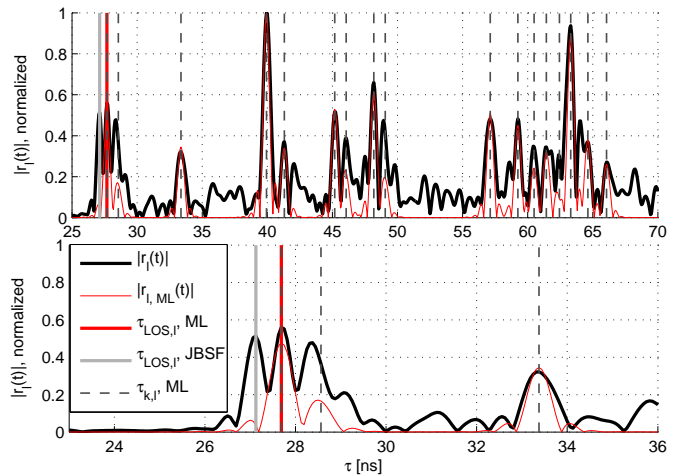}
\caption{Exemplary ranging results for BS 4 and position $l=120$ (c.f. Fig. \ref{fig:scen}), using $\Tp = \vu{0.5}{ns}$. The lower plot is a close-up. The black line denotes the received signal, while the red signal illustrates the reconstruction using the ML channel estimator. The latter also provides estimates for the MPC arrival times $\tau_{k,l}$ (grey dashed lines), where the ML range estimate (in red) is the minimum of these. Here, JBSF (light grey) outperforms ML due to path overlap. The simplified implementation of the ML channel estimator assumes a separable channel and only detects the larger second component.}
\label{fig:range}
\end{figure}

\subsection{Jump-back Search-forward (JBSF) Ranging}
If real-time localization is desired, a channel estimation like \eqref{eq:clean} might not be feasible. JBSF is a threshold-based ranging algorithm that also works for low-complexity receiver structures like energy detectors \cite{Sahinoglu2008}. Backwards from the maximum of the received signal at delay $\taumax$, a search-back window of length $\tausb$ is defined. The JBSF range estimate is the minimum delay within this window that exceeds a certain threshold
\begin{equation}
   \tauLOSJBSF = \min \{ \hat{\tau} : \taumax-\tausb  \leq \hat{\tau} \leq \taumax, |r_\ell(\hat{\tau})| \geq A_\xi \}.
  \label{eq:JBSFrange}
\end{equation}
The choice of the threshold $A_\xi$ is usually based on the noise level.

An illustrative example for the channel estimation and ranging algorithms is given in Fig. \ref{fig:range}. In this case, the LOS path is not separable from the rest of the channel. Since the amplitude of the second arriving MPC is larger, it is detected in an earlier iteration of the ML estimator. Due to the separable channel assumption, the ML estimator is constrained not to estimate any MPCs within one $\Tp$ of already detected ones, hence the LOS path is not detected. However, the LOS component is above the threshold of JBSF and within its search-back window, so JBSF can outperform the ML estimator in this case.

\section{Tracking and Data Association}
\label{sec:tracking}

The conventional tracking schemes based on ranging and the MINT approach are both implemented in this work using extended Kalman filters (EKFs) \cite{Simon2006,Bar-Shalom1988} employing linearized observation equations for the estimated path delays. As in \cite{MeissnerIWSSIP2012}, we use a simple constant-velocity state-space model of the moving agent node. Stacking the position $\vm{p}=[p_x, p_y]^T$ and velocity $\vm{v}=[v_x, v_y]^T$ in the state vector $\vm{x}=[\vm{p}^T, \vm{v}^T]^T$ the model reads
\begin{align}
    \vm{x}_{\ell+1} = & \vm{F}\vm{x}_{\ell} + \vm{G}\vm{n}_{\mathrm{a},\ell} \\
    = &
     \begin{bmatrix}
       1 & 0 & \Delta T & 0 \\
       0 & 1 & 0 & \Delta T \\
       0 & 0 & 1 & 0\\\
       0 & 0 & 0 & 1
     \end{bmatrix}
    \vm{x}_{\ell} +
    \begin{bmatrix}
      \frac{\Delta T^2}{2} & 0 \\
      0 & \frac{\Delta T^2}{2} \\
      \Delta T & 0 \\
      0 & \Delta T \\
    \end{bmatrix}
 \vm{n}_{\mathrm{a},\ell}. \notag \label{eq:ssmodel}
\end{align}
Here, $\Delta T$ is the update interval and $\vm{n}_{\mathrm{a},\ell}$ denotes the driving process noise, which allows for motion at non-constant velocity.
The measurement inputs for the conventional tracking scheme using EKFs are either the ML or JBSF ranging estimates. Using \eqref{eq:MLrange} or \eqref{eq:JBSFrange}, ranging is performed to each of the four BSs. The entries of the vector of distance estimates are obtained as $z_{\ell}^{(i)}=c \tauLOSMLi$ or $z_{\ell}^{(i)}=c \tauLOSJBSFi$, where $i$ is the BS index.

\subsection{MINT Using EKF with Data Association (DA)}
\label{sec:MINTDA}

For the MINT scheme implemented using an EKF, the MPC estimation in \eqref{eq:clean} only results in an unordered set of distance estimates, i.e. $\Z{\ell}^{(i)}=c\cdot\{ \hat{\tau}_{k,\ell}^{(i)} \}$, where the cardinality of the set is $|\Z{\ell}^{(i)}|=\hat{K}_\ell^{(i)}$. To obtain a suitable input for the observation equations of the EKF, a DA scheme is necessary. As the association to the BS is known, we only consider the DA for one BS and drop the index $i$. After the DA, all associated distance estimates are stacked into one measurement vector.

The employed DA approach is the same as used in \cite{MeissnerIWSSIP2012} and is just shortly re-sketched here. We introduce a set $\C{\ell}$ of correspondence variables \cite{Thrun2006},
whose $n$-th entry $c_{\ell,n}$ represents the association of the $n$-th entry of $\Z{\ell}$ and is defined as
\begin{equation}
  c_{\ell,n} = \begin{cases}
              j, & \text{if $z_{\ell,n}$ corresponds to VA $\va{j}$ } (j=1,\dots,N_{\mathrm{VA}})\\
              0, & \text{if $z_{\ell,n}$ corresponds to clutter.}
            \end{cases}
            \label{eq:corrvar}
\end{equation}

Using the predicted position of the EKF, $\vm{p}_\ell^-$, the set of expected visible VAs $\VA{\ell}$ can be computed using \eqref{eq:fvis} and \eqref{eq:VAsl}. From this, the set of expected path lengths, denoted as $\D{\ell}$, can be obtained by calculating the Euclidean distances $d(\va{j}, \vm{p}_\ell^-)$ between all VAs $\va{j} \in \VA{\ell}$ and the position $\vm{p}_\ell^-$. The task of the DA is now to match the two sets $\Z{\ell}$ and $\D{\ell}$ in an optimum way.

The approach used is optimal sub-pattern assignment \cite{Schuhmacher2008}. Noting that the cardinality of $\D{\ell}$ is $K_\ell$ (c.f. \eqref{eq:VAsl}) and assuming that $\hat{K}_\ell \geq K_\ell$, which can always be ensured by filling up $\Z{\ell}$ with dummy clutter, we search the sub-pattern of $\Z{\ell}$ that fits best to the set $\D{\ell}$. Using a distance function $d^{(\cutoff)}(\cdot)$, that simply saturates at the cut-off distance $\cutoff$, this is expressed as
\begin{equation}
  \boldsymbol{\pi}_{\mathrm{opt}} = \arg \min_{\boldsymbol{\pi} \in \Pi_{\hat{K}_\ell}} \sum_{i=1}^{K_{\ell}} d^{(\cutoff)} (d_{\ell,i},
z_{\ell,{\pi}_i} ).
\label{eq:piopt}
\end{equation}
Here, $\Pi_{\hat{K}_\ell}$ is the set of permutations of positive integers up to $\hat{K}_\ell$ and the vector $\boldsymbol{\pi}_{\text{opt}}$ holds the permutation with the minimum cumulative distance in the sense of the cutoff distance function $d^{(\cutoff)}(\cdot)$. Hence, the first $K_\ell$ entries of $\boldsymbol{\pi}_{\text{opt}}$ contain the indices of those entries of $\Z{\ell}$ that have been optimally assigned to $\D{\ell}$.

As a last step in the DA process, we reject those measurements for which $d^{(\cutoff)}(d_{\ell,i}, z_{\ell,\pi_{\text{opt},i}})=\cutoff$, i.e. those that have been assigned at a distance greater or equal than the cutoff distance. This is done to have a mechanism to tune the DA to the uncertainty in the MPC delay estimates. As UWB signals are used, if an MPC is detectable at all, its delay will have a relatively small uncertainty that depends on the used pulse duration $\Tp$ and the inaccuracies of the floor plan. Hence, the correspondence variables in \eqref{eq:corrvar} are obtained as
\begin{equation}
  c_{\ell,n} = \begin{cases}
              j, & \text{ if ${\pi}_j = n$ and $d^{(\cutoff)} (d_{\ell,j}, z_{\ell,\pi_{\text{opt},n}}) < \cutoff$} \\
              0, & \text{ else.}
            \end{cases}
            \label{eq:DA}
\end{equation}
The optimal sub-pattern assignment can be efficiently implemented using the Munkres algorithm \cite{Munkres1957}.

\subsection{Genie-aided Data Association (GADA) for Evaluation}
The DA discussed above depends crucially on the EKF's predicted position $\vm{p}_\ell^-$. To enable an evaluation of the localization performance where the errors due to false DA are mostly eliminated, we use a so-called genie-aided version of the DA (GADA). The GADA relies on the true position $\vm{p}_\ell$ for performing the DA, i.e. in the calculation of the set $\D{\ell}$. This can be used to benchmark the usefulness of the MPC delay estimates for localization using MINT.


\section{Results}
\label{sec:results}

\begin{table*}[!ht]
\renewcommand{\arraystretch}{1.3}
\caption{Parameters and results for the performance evaluation. For the performance measures, a pair of values is given where the first one indicates the scenario without the obstruction and the second one with the obstruction.}
\label{tab:results}
\centering
\begin{tabular}{c|c|c|c|c|c}
  & $\Tp=\vu{0.2}{ns}$ & $\Tp=\vu{0.5}{ns}$ & $\Tp=\vu{1}{ns}$ & $\Tp=\vu{2}{ns}$ & $\Tp=\vu{4}{ns}$ \\
  \hline
  \hline
  Process noise variance $\sigmaa$ $[\text{m}^2/\text{s}^4]$ & \multicolumn{5}{c}{$\sigmaa = (\frac{\vxmax}{3 \Delta T})^2$, covariance matrix $\vm{Q}=\sigmaa\vm{GG}^T$, with $\vxmax=\vu{1.5}{m/s}$} \\
   \hline
  EKF measurement noise variance $\sigmaz$ $[\text{m}^2]$ & $0.01$ & $0.01$ & $0.04$  & $0.04$ & $0.09$\\
   \hline
  DA cutoff distance $\cutoff$ (MINT) $[\text{m}]$ & $0.3$ & $0.3$ & $0.5$ & $0.5$ & $0.6$ \\
  \hline
  JBSF SB-window $[\text{ns}]$ & \multicolumn{5}{c}{$100$} \\
  \hline
  JBSF threshold $\xi$ $[1]$ & $0.4$ & $0.4$ & $0.3$ & $0.3$ & $0.3$ \\
  \hline
  \hline
  Mean number of assoc. MPCs (EKF-GADA) [1] & $18.2$ / $17.6$ & $18.6$ / $18.2$ &  $18.3$ / $17.9$ & $15.0$ / $13.9$ & $11.9$ / $10.2$\\
  \hline
  Mean number of assoc. MPCs (EKF-DA) [1] & $18.1$ / $17.5$ & $18.5$ / $18.2$ & $18.2$ / $17.7$ & $15.1$ / $14.0$ & $11.8$ / $10.1$\\
  \hline
  \hline
  RMS position error MINT (EKF-GADA) $[\text{m}]$ & $0.03$ / $0.032$ & $0.039$ / $0.041$ & $0.06$ / $0.069$ & $0.076$ / $0.083$ & $0.117$ / $0.136$\\
  \hline
  RMS position error MINT (EKF-DA) $[\text{m}]$ & $0.073$ / $0.087$ & $0.067$ / $0.065$ &  $0.118$ / $0.197$ & $0.122$ / $0.149$ & $0.207$ / $0.304$\\
  \hline
  RMS position error EKF, ML ranging $[\text{m}]$ & $0.05$ / $0.052$ & $0.058$ / $0.065$ & $0.079$ / $0.113$ &  $0.129$ / $0.149$ & $0.353$ / $0.396$\\
  \hline
  RMS position error EKF, JBSF ranging $[\text{m}]$ & $0.113$ / $0.15$ & $0.351$ / $0.418$ & $0.243$ / $0.297$ & $0.249$ / $0.34$ & $0.406$ / $0.714$\\
  \hline
  \hline
  average HDOP MINT (EKF-GADA) $[1]$ & $0.41$ / $0.39$ & $0.41$ / $0.38$ & $0.39$ / $0.39$ &  $0.43$ / $0.43$ & $0.47$ / $0.51$\\
  \hline
  average HDOP MINT (EKF-DA) $[1]$ & $0.67$ / $0.73$ & $0.56$ / $0.53$ & $0.56$ / $0.65$ & $0.58$ / $0.61$ & $0.68$ / $0.76$\\
  \hline
  average HDOP EKF, ML ranging $[1]$ & $1.19$ / $1.21$ & $1.23$ / $1.2$ & $1.35$ / $1.23$ & $1.09$ / $0.96$ & $1.04$ / $0.8$\\
  \hline
  average HDOP EKF, JBSF ranging $[1]$ & $1.23$ / $1.28$ & $1.27$ / $1.52$ & $1.56$ / $1.49$ & $1.28$ / $1.24$ & $1.03$ / $1.14$\\
  \hline
\end{tabular}
\end{table*}

\begin{figure}[t]
\centering
\includegraphics[width=1\columnwidth,keepaspectratio=true]{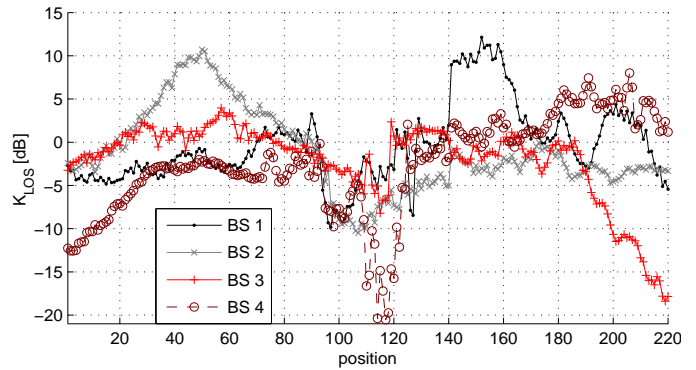}
\caption{Instantaneous K-factor with respect to the LOS component $K_\text{LOS}$ for all BSs over the trajectory using a pulse duration $\Tp=\vu{0.5}{ns}$.}
\label{fig:KLOS}
\end{figure}

This section illustrates the accuracy and robustness of the proposed indoor tracking algorithms for the scenario in Fig. \ref{fig:scen}. A first rough evaluation of the propagation conditions can be obtained by evaluating the K-factor with respect to the LOS component, i.e. the ratio of its energy to the energy of the rest of the CIR. It is estimated using \eqref{eq:clean} and \eqref{eq:alpha}, where only the LOS component is subtracted from $r_\ell(t)$, i.e.
\begin{equation}
  \hat{K}_{\text{LOS},\ell} = \frac{ |  \hat{\alpha}_{\text{LOS},\ell} |^2 }{ \int_T  | r_\ell(t) - \hat{\alpha}_{\text{LOS},\ell} s(t-\hat{\tau}_{\text{LOS},\ell})|^2 dt }
\end{equation}
where we note that the pulse $s(t)$ is energy-normalized.

Fig. \ref{fig:KLOS} shows  $\hat{K}_{\text{LOS},\ell}$ for all BSs over the trajectory, estimated using the original measurements (without the obstruction) with $\Tp=\vu{0.5}{ns}$. Some geometric influences, like the drop in $K_{\text{LOS}}$ in the NLOS regions of BSs 3 and 4 or the general distance dependence can be easily observed. Other causes of large variations are not immediately intuitive and require more careful inspection of the local propagation conditions. For example, the drop of $K_{\text{LOS}}$ for BS 1 around position 90 is due to strong scattered components from metal pillars within the lower windows, c.f. Fig. \ref{fig:scen}.

\subsection{Evaluation Setup}

The ``obstruction'' introduced in Fig. \ref{fig:scen} is meant to model a change in the propagation environment, e.g. a group of people. As it causes NLOS regions for all BSs, its influence on the tracking performance is interesting. MINT is not aware of this obstruction in its floor plan knowledge. The original channel measurements where performed without this obstruction, hence they need to be modified. For this, we calculate the positions which are now in NLOS due to the changed conditions. At these positions, we reduce $K_\text{LOS}$ by a factor of \vu{10}{dB}, which is done by subtracting the LOS component and adding it again with a suitable amplitude and its original phase. To enable a fair comparison with MINT, we also change the amplitudes of all the MPCs affected by the change in visibility conditions. We do so by reducing their SINR \eqref{eq:sinr} also by \vu{10}{dB}. Note that this factor leaves the LOS component still detectable in most cases, while for the later arriving MPCs it causes an effective loss of the respective component

\subsection{Choice of Parameters}
\label{sec:results_param}

\paragraph{Measurement parameters}
We perform the performance evaluation for several exemplary signal bandwidths, i.e. the pulse duration of the raised cosine pulse is chosen as $\Tp \in \{ \vu{0.2}{ns}, \vu{0.5}{ns}, \vu{1}{ns}, \vu{2}{ns},\vu{4}{ns} \}$. The corresponding \vu{-3}{dB} bandwidths are $\{ \vu{5}{GHz}, \vu{2}{GHz}, \vu{1}{GHz}, \vu{0.5}{GHz},\vu{0.25}{GHz} \}$. Using a roll-off factor of $\beta_\text{R}=0.5$, the center frequency is always $f_c=\vu{7}{GHz}$, except for $\Tp=\vu{0.2}{ns}$ which corresponds to the full measurement bandwidth, hence the overall center frequency $f_c=\vu{6.85}{GHz}$ has to be used. Table \ref{tab:results} contains an overview of all relevant parameters for these setups.

\paragraph{EKF parameters}
The EKF's measurement noise covariance matrix is $\vm{R} = \sigmaz \vm{I}$, where $\vm{I}$ is the identity matrix. For the variance $\sigmaz$, we choose values based on the expected ranging uncertainty, i.e. it is slightly increased with the pulse duration $\Tp$. The same consideration applies for the DA cutoff distance $\cutoff$ used in \eqref{eq:piopt} and \eqref{eq:DA}. Table \ref{tab:results} lists the values used in the different measurement parameter setups.

The choice of the process noise variance $\sigmaa$ depends on the maximum velocity in e.g. the $x$-direction, denoted as $\vxmax$. The process noise, assumed to be Gaussian, should allow movement at this velocity. Hence, we select the $3\sigma$ point of the velocity noise as $\vxmax$ and obtain for the corresponding process noise  variance in the acceleration domain
\begin{equation}
  \sigmaa = \left( \frac{\vxmax}{3\Delta T} \right)^2.
\end{equation}
In this paper, we choose to model pedestrian motion and set $\vxmax = \vu{1.5}{m/s}$. The update interval is $\Delta T=\vu{0.1}{s}$, resulting in a magnitude of the velocity in $x$- or $y$-direction of $\vu{1}{m/s}$ along straight segments of the trajectory.
The conventional tracking schemes and MINT always use the same $\sigmaz$ and $\sigmaa$ for every pulse duration considered.

\paragraph{Initialization}
All EKFs are initialized with the correct position and zero velocity. We note that this is a more challenging problem for MINT than for the conventional schemes, since the DA depends on the estimated position. However, we have addressed this problem for a different implementation of MINT in \cite{MeissnerIPIN2010} using a Gaussian-sum filter \cite{Alspach1972}. In this approach, a bank of parallel weighted Kalman filters (KFs) is used to represent multiple initial position hypotheses. After convergence, the KFs with small weights can be discarded. This can also be applied for the MINT implementation in this paper.

\begin{figure}[t]
\centering
\includegraphics[width=1\columnwidth,keepaspectratio=true]{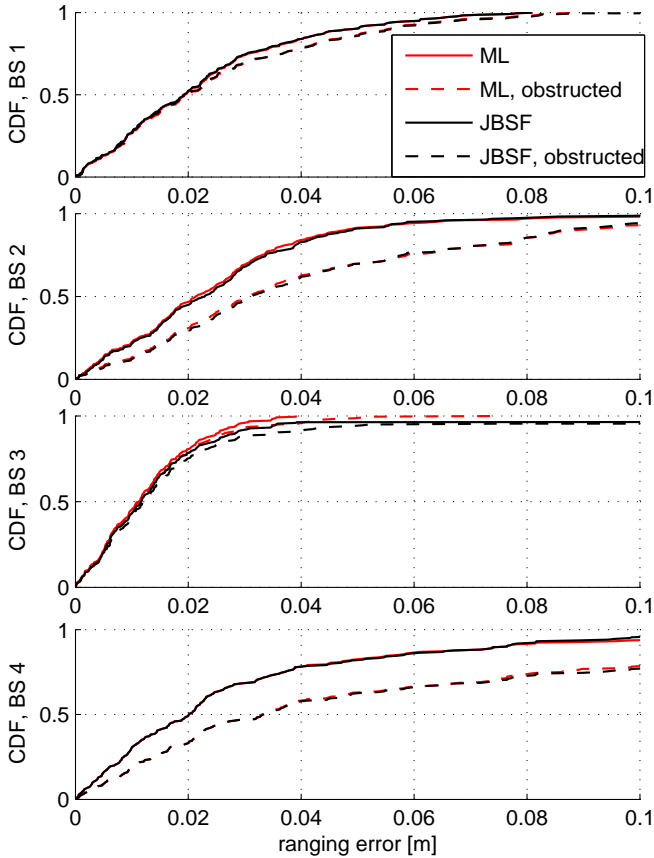}
\caption{CDFs of the ranging error for ML (red) and JBSF (black) with and without the obstruction for all BSs and $\Tp=\vu{1}{ns}$.}
\label{fig:rangingCDFs}
\end{figure}

\paragraph{Channel estimation and ranging}
The choice of $\hat{K}_\ell$ in \eqref{eq:clean}, i.e. the number of extracted components, is based on the average number of expected visible VAs in the localization environment. For the performance evaluations, we take into account VAs up to order two and fix $\hat{K}_\ell=\hat{K}=20$. However, we define a threshold between the noise floor and the maximum amplitude of the received signal, below which no further MPCs are extracted. The noise signal $\hat{w}_\ell(t)$ is estimated from the pre-LOS part of $r_\ell(t)$. Denoting the time average of a signal with $\langle\cdot\rangle$, the threshold is obtained as \cite{MeissnerICUWB2011}
\begin{equation}
   A_{\gamma, \ell} = \gamma\cdot (\max\{ |r_\ell(t)| \} -   \langle | \hat{w}_\ell(t)  | \rangle  ) + \langle | \hat{w}_\ell(t) | \rangle.
  \label{eq:threshold}
\end{equation}
In this paper, $\gamma=0.1$ is used.

The JBSF ranging threshold $A_\xi$ in \eqref{eq:JBSFrange} is also chosen using a relative threshold $\xi$, in the same manner as $\gamma$ in \eqref{eq:threshold}. For every pulse duration used, we selected a threshold at which ranging works well in the unobstructed scenario, c.f. Table \ref{tab:results}. The length of the search-back window $\tausb$ is chosen for each BS as the maximum distance between the BS and any first-order VA, expressed in ns. This corresponds to the maximum delay difference of the LOS component and any first-order reflection. The reasoning behind this is the assumption that if the LOS component is not the maximum itself, then the maximum will be among the first-order MPCs. A value of \vu{100}{ns} is selected for the geometry in Fig. \ref{fig:scen}. It approximately corresponds to the excess delay of the reflection path on the far right end of the corridor.

\begin{figure}[t]
\begin{center}
\subfloat[$\Tp=\vu{0.2}{ns}$]{
     \includegraphics[width=1.00\columnwidth]{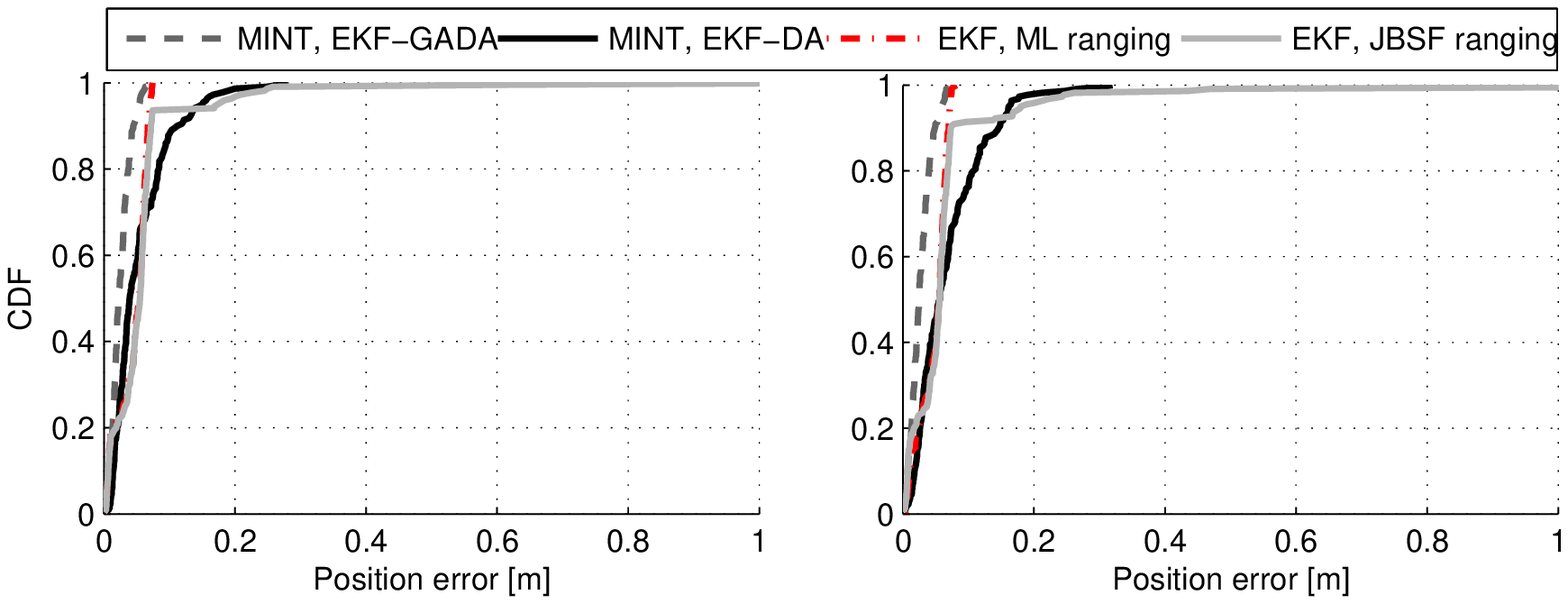}
     \label{fig:CDF:02} 
  }
  \\
  \subfloat[$\Tp=\vu{0.5}{ns}$]{
     \includegraphics[width=1.00\columnwidth]{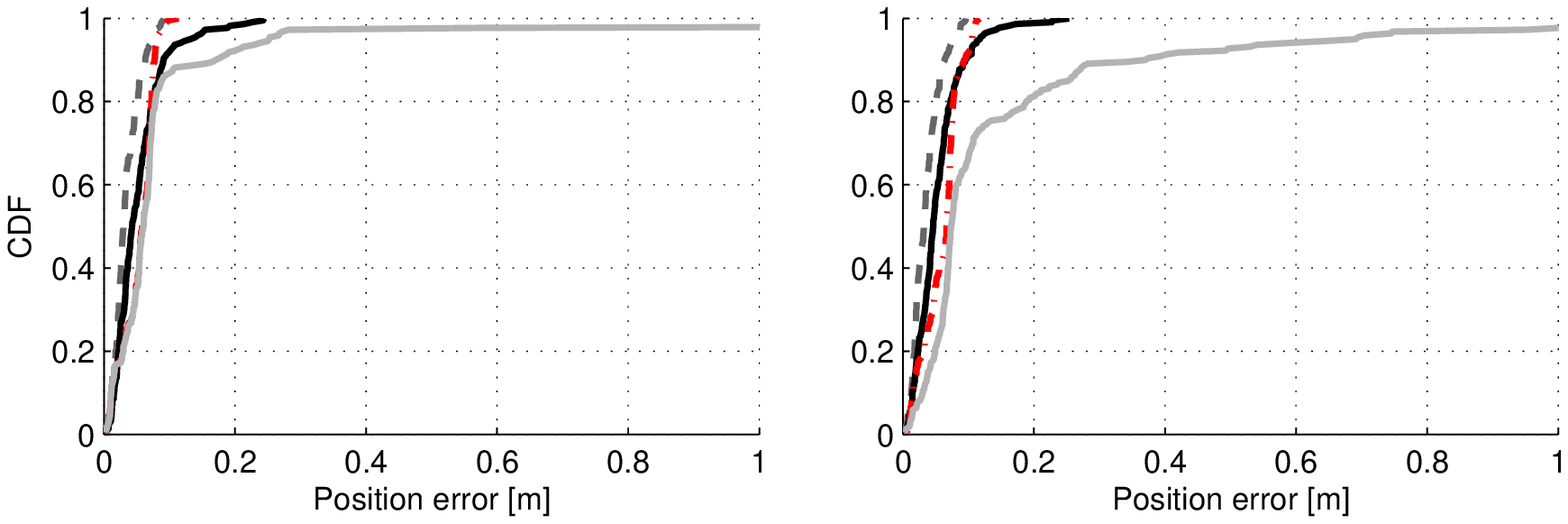}
     \label{fig:CDF:05} 
  }
  \\
  \subfloat[$\Tp=\vu{1}{ns}$]{
     \includegraphics[width=1.00\columnwidth]{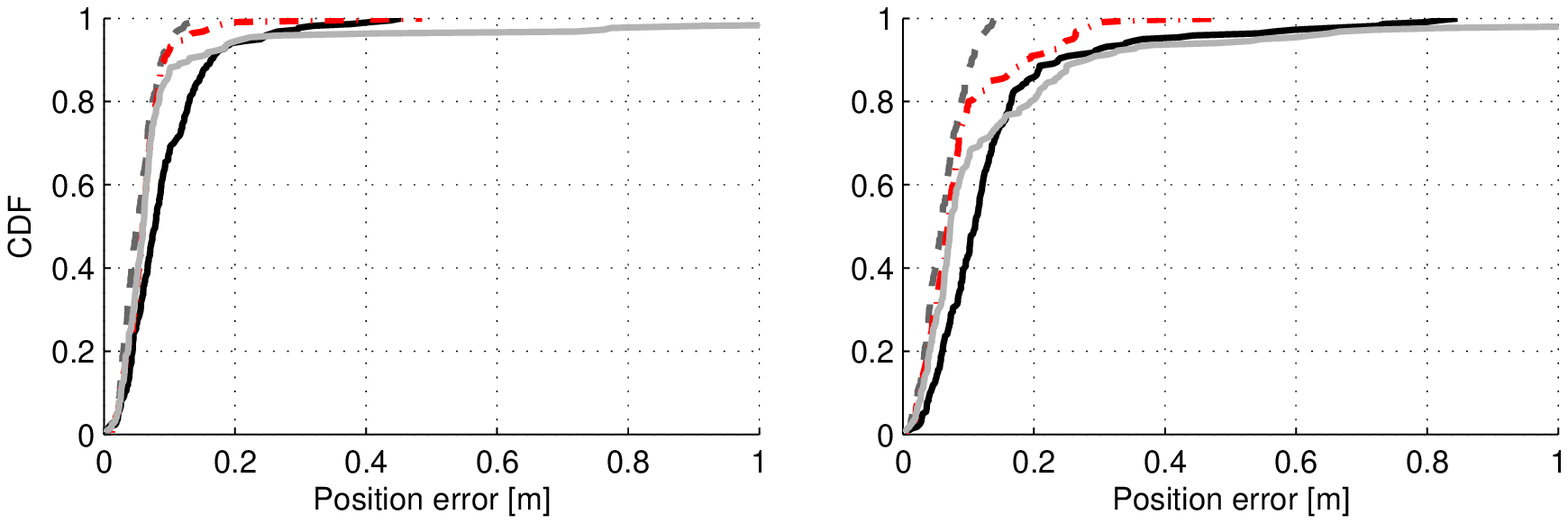}
     \label{fig:CDF:1} 
  }
  \\
  \subfloat[$\Tp=\vu{2}{ns}$]{
     \includegraphics[width=1.00\columnwidth]{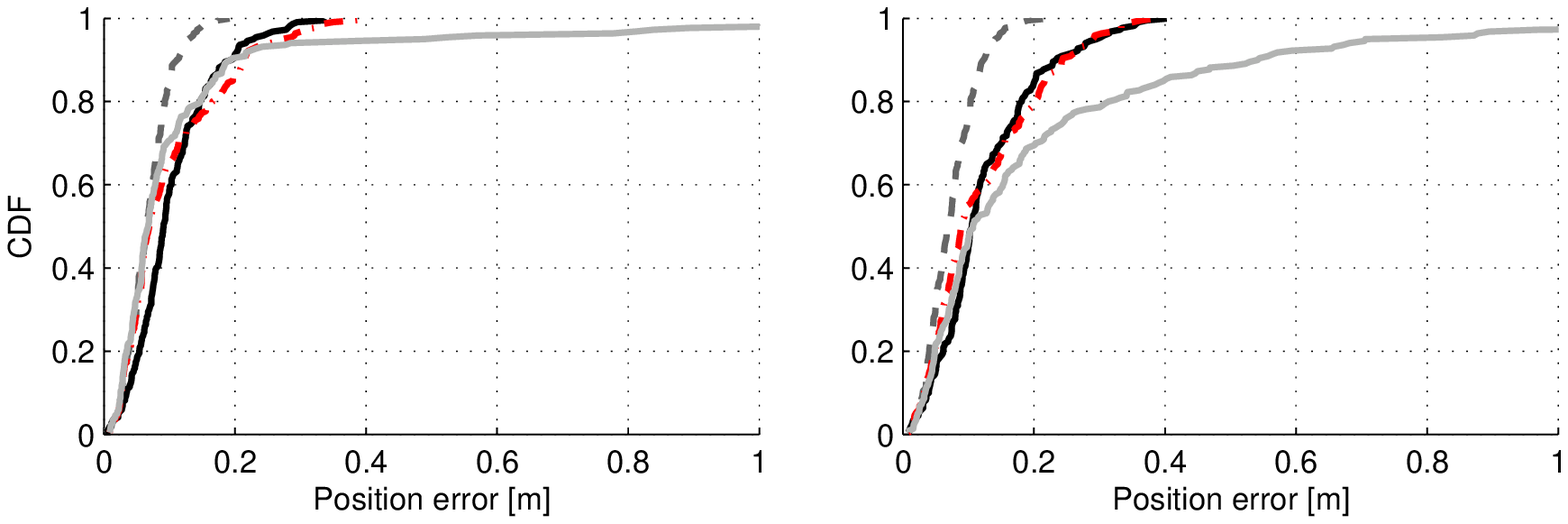}
     \label{fig:CDF:2} 
  }
  \\
  \subfloat[$\Tp=\vu{4}{ns}$]{
     \includegraphics[width=1.00\columnwidth]{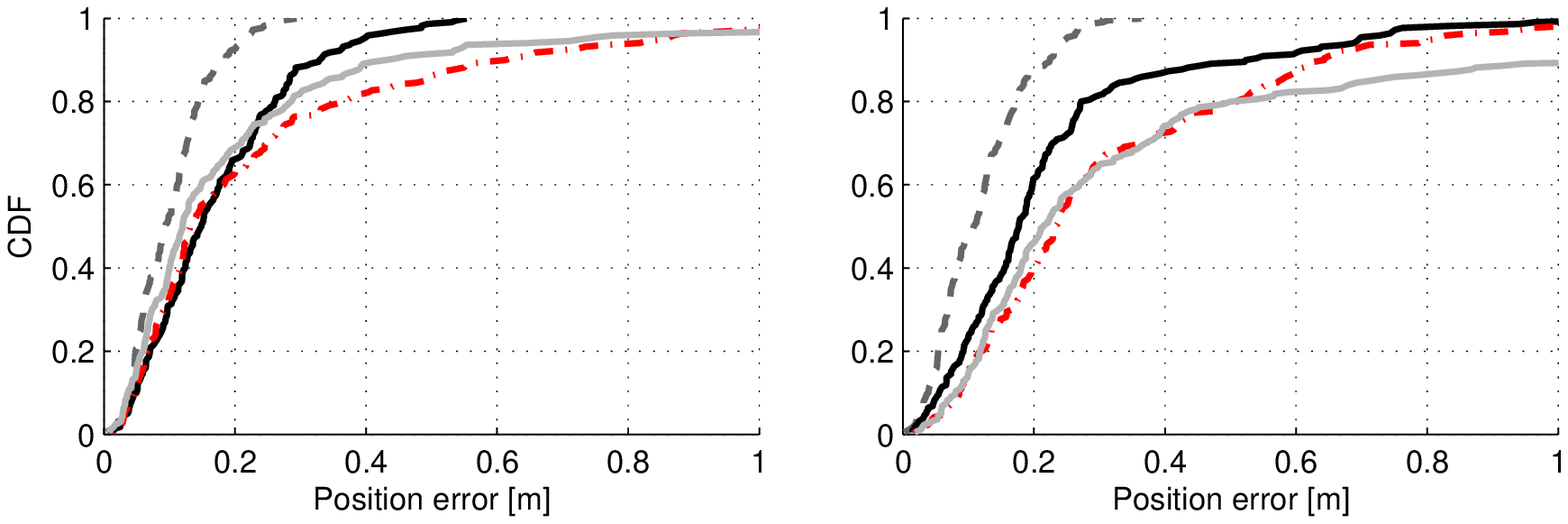}
     \label{fig:CDF:4} 
  }
  \end{center}
\caption{CDFs of the position error over the trajectory for all considered pulse durations $\Tp$. The left column shows runs without the obstruction (c.f. Fig. \ref{fig:scen}), in the right column, signals have been modified according to the obstruction.}
\label{fig:cdfs}
\end{figure}

\subsection{Ranging Performance Results}
A comparison of the performance of ML and JBSF ranging algorithms in terms of the ranging error CDF is shown in Fig. \ref{fig:rangingCDFs} with and without the obstruction. A pulse duration of $\Tp=\vu{1}{ns}$ has been used for this. The most striking result is that JBSF closely approaches ML ranging for all cases, at a much lower computational complexity. One reason for the good performance of JBSF is the rather low noise level in the VNA measurements. However, when looking at the instantaneous ranging errors over the trajectory (not shown), some isolated outliers of JBSF can be observed. The implications for the localization performance will be discussed in the next subsection.

\subsection{Localization and Tracking Performance Results}

Table \ref{tab:results} contains detailed performance results in terms of RMS position error, the mean number of associated MPCs for MINT and the average horizontal dilution of precision (HDOP). The latter value is calculated as the average of the instantaneous HDOP values over the trajectory. These are the ratio of instantaneous position error and RMS ranging error \cite{Sahinoglu2008}, hence they provide snap-shots of the quality of the localization geometry.
For MINT, the RMS ranging error is computed from those range estimates that have been associated to VAs by the DA procedure.

Looking at the RMS position error in Table \ref{tab:results}
obtained using both ranging methods reveals that the EKF using ML ranging still clearly outperforms the one using JBSF. This behavior would not be suspected by looking at the ranging CDFs alone. The few isolated outliers of JBSF mentioned above cause the gap in localization performance. As this subsection shows, MINT does not show such robustness issues with respect to ranging outages.

Fig. \ref{fig:cdfs} contains the position error CDFs for all considered pulse durations. MINT using EKF-GADA always shows the best performance of all tracking algorithms. Although this is just an upper bound for this implementation of MINT, it shows the potential gain of exploiting multipath propagation. This is not only true for the achieved accuracy, but also for the robustness with respect to difficult propagation conditions, since MINT using EKF-GADA shows almost no performance loss for the obstructed scenario. Also with realistic DA, MINT is able to overcome impairments caused by the obstruction better than the conventional approaches. The conventional approaches show significant performance impairments, especially when the pulse duration is increased.

Only for pulse durations of $\Tp=\vu{0.2}{ns}$ and $\Tp=\vu{1}{ns}$, MINT with DA can not provide a performance gain. In the first case, the channel estimation can resolve very closely spaced MPCs that are challenging for the DA due to their clustered structure. In the second case, interference by pulse overlap situations causes the detection of several MPCs to fail. Since neither the channel estimation \eqref{eq:clean} nor the DA as proposed here can deal with path overlap situations, this is identified as an important topic for future research.

An unexpected result is obtained for $\Tp=\vu{4}{ns}$, which, corresponding to a \vu{-3}{dB} bandwidth of \vu{250}{MHz}, is no longer UWB by definition \cite{Molisch2009}. Although MINT generally benefits from a large bandwidth, it can yield the largest performance and robustness gain here. The conventional tracking schemes show outliers due to very large ranging biases, while MINT still exploits information from some MPCs, although their number decreases significantly (c.f. Table \ref{tab:results}). The mentioned large ranging biases are also the reason for the lower HDOP in the obstructed scenario in these parameter setups. Furthermore, it should be noted that in terms of the RMS position error, MINT using the EKF-DA always outperforms the conventional EKF approach using JBSF ranging.

\begin{figure}[t]
\begin{center}
  \includegraphics[width=1\columnwidth]{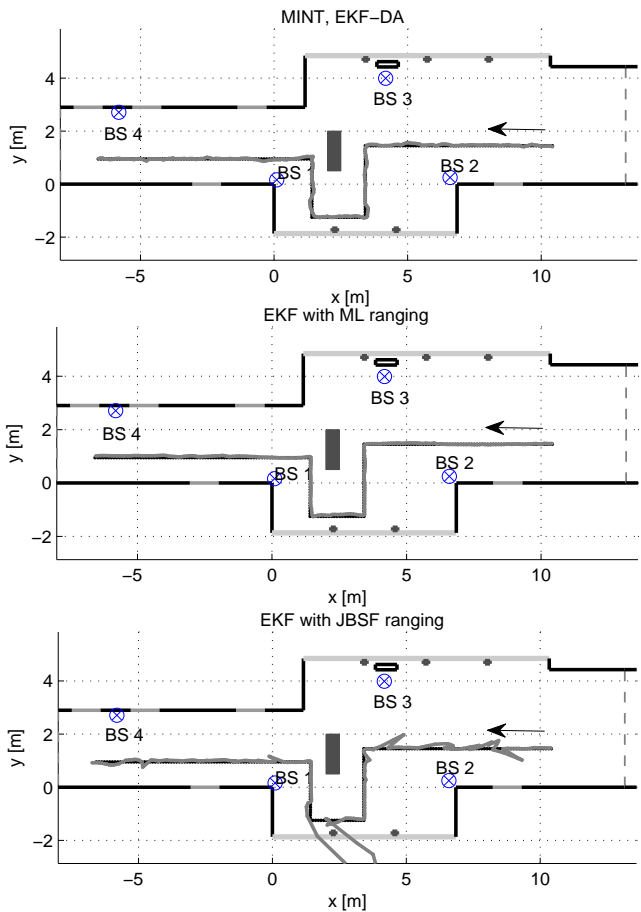}
  \caption{Tracking results for the NLOS scenario and $\Tp=\vu{0.5}{ns}$. The three plots show the estimated trajectories using MINT with the EKF and data association, and EKFs using ML and JBSF ranging, respectively.}
     \label{fig:resgeom:traj} 
\end{center}
\end{figure}

\begin{figure}[t]
\begin{center}
     \includegraphics[width=.95\columnwidth]{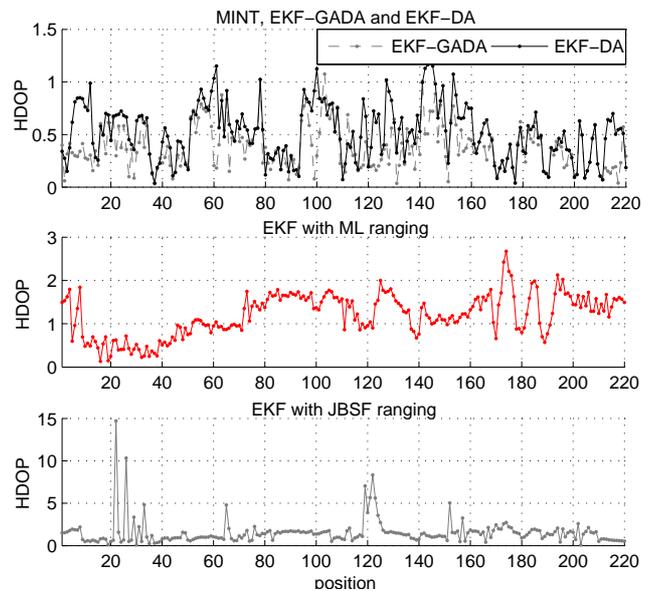}
       \caption{Tracking results for the NLOS scenario and $\Tp=\vu{0.5}{ns}$. The plots illustrate the HDOP for the tracking algorithms used.}
       \label{fig:resgeom:hdop} 
\end{center}
\end{figure}

An illustrative example for $\Tp=\vu{0.5}{ns}$ and the NLOS scenario  is shown in Figs. \ref{fig:resgeom:traj} and \ref{fig:resgeom:hdop}. The estimated trajectories of the agent node are displayed in Fig. \ref{fig:resgeom:traj}. While MINT and the EKF using ML range estimates can both follow the trajectory well, the EKF employing JBSF ranging is severely impaired by the NLOS regions of BSs 1 and 4 (at the start of the trajectory), BS 4 (around position 120, c.f. Fig. \ref{fig:scen}) and BS 3 and 2 (near the end of the trajectory). As Fig. \ref{fig:resgeom:hdop} confirms, in the first two of these NLOS regions, rather large values of the instantaneous HDOP occur. This highlights the bad geometry for conventional schemes when there are ranging outages. For MINT, most of the HDOP values along the trajectory are below one, indicating excellent geometry for localization. This is due to the large set of virtual BSs spanned by the VAs (c.f. \eqref{eq:efim1} and \eqref{eq:RDM}).

\section{Conclusion and Future Work}

We have presented a comparison of conventional indoor UWB ranging and tracking schemes and an implementation of the proposed MINT approach using data from a measurement campaign in a representative indoor environment. The influence of NLOS situations on the accuracy and robustness of these systems has been investigated. Results confirm the potentials of MINT: Using floor plan information, multipath propagation can be efficiently used to obtain position-relevant information in situations where conventional approaches fail. Even at comparably low bandwidths, MINT shows a clear gain in localization performance.

Ongoing and future work includes the modeling of floor plan uncertainties and the combination of localization and channel estimation. The latter can be used to reduce the amount of clutter in the extracted path delay estimates. Also, other implementations of MINT, e.g. using soft classification in the data association, are within the scope of our work.

\section*{Acknowledgment}
The authors thank Daniel Arnitz and Thomas Gigl for their invaluable help in the channel measurement campaign.
This work was partly supported by the Austrian Science Fund (FWF) within the National Research Network SISE project S10610, and by the Austria Research Promotion Agency (FFG) within KIRAS PL3, grant no. 832335 ``LOBSTER''.



 \bibliographystyle{IEEEtran}
 \bibliography{IEEEabrv,../../PhDReading}
%
%

\end{document}